\documentstyle[epsfig,preprint,aps]{revtex}

\input prepictex
\input pictex
\input postpictex

\makeatletter
\def\references{%
\ifpreprintsty
\bigskip
\hbox to\hsize{\hss\large \refname\hss}%
\else
\vskip24pt
\hrule width\hsize\relax
\vskip 1.6cm
\fi
\list{\@biblabel{\arabic{enumiv}}}%
{\labelwidth\WidestRefLabelThusFar  \labelsep4pt %
\leftmargin\labelwidth %
\advance\leftmargin\labelsep %
\ifdim\baselinestretch pt>1 pt %
\parsep  4pt\relax %
\else %
\parsep  0pt\relax %
\fi
\itemsep\parsep %
\usecounter{enumiv}%
\let\p@enumiv\@empty
\def\theenumiv{\arabic{enumiv}}%
}%
\let\newblock\relax %
\sloppy\clubpenalty4000\widowpenalty4000
\sfcode`\.=1000\relax
\ifpreprintsty\else\small\fi
}
\makeatother

\def\h{\frac{1}{2}}
\def\t{\frac{1}{3}}
\def\q{\frac{1}{4}}

\draft
\preprint{}
\begin{document}
\title{\Large\bf Cosmic Axion\footnote{Talk presented
at The 2$^{nd}$ Int. Workshop on Gravitation and Astrophysics,
ICRR, Univ. of Tokyo, Nov. 17--19, 1997}}
\author{Jihn E. Kim} 
\baselineskip=0.5cm
\address{Lyman Laboratory of Physics, Harvard University,
Cambridge, MA 02138, and\\
Department of Physics, Seoul National University,
Seoul 151-742, Korea \footnote{Permanent address}}
\maketitle

\begin{abstract} 
I review the axionic solution of the strong CP problem
and current status of the cosmic axion search.
\end{abstract}

\def\h{\frac{1}{2}}
\def\t{\frac{1}{3}}
\def\q{\frac{1}{4}}

\section{Introduction}

Quantum chromodynamics before 1975 considered the following Lagrangian
\begin{equation}
{\cal L}=-{1\over 2g^2}{\rm Tr}F_{\mu\nu}\tilde F^{\mu\nu}
+\bar q(iD^\mu\gamma_\mu-M)q.
\end{equation}
where $M$ is the diagonal, $\gamma_5$-free, real quark mass matrix.
But after 1975, the following term is known to be prensent
in general in a world without a massless quark,
\begin{equation}
+{\bar\theta\over 16\pi^2}{\rm Tr}F_{\mu\nu}\tilde F^{\mu\nu}
\end{equation}
Since this $\bar\theta$ term violates CP invariance, the upper
bound of the neutron electric dipole moment puts a strong
constraint on the magnitude of $\bar\theta$,
$|\bar\theta|<10^{-9}$. The smallness of $\bar\theta$
has led to the strong CP problem, $\lq\lq$Why is $\bar\theta$
so small?"~\cite{strcp}

We know that many small parameters in physics have led to new
ideas, in most cases leading to new symmetries. For example,
$M_W/M_P\ll 1$ has led to supersymmetry, $m_{u,d}\ll 1$ GeV has 
led to $SU(2)_L\times SU(2)_R$ chiral symmetry, etc. For the
strong CP problem, the nicest solution is the very light
axion resulting from the Peccei-Quinn symmetry~\cite{pq}.

\section{The Axion Solution}

The reason that the axion solves the strong CP problem is the
following. This argument is due to Ref.~\cite{vw}. In the
axion solution, $\bar\theta$ is a dynamical field, but for
a moment let us treat it as a parameter (or coupling constant).
The partition function in the Euclidian space after integrating 
out the quark fields is
\begin{equation}
e^{-V[\bar\theta]}\equiv\int [dA_\mu]\prod_i{\rm Det}(D^\mu\gamma_\mu
+m_i)\exp[-\int d^4x({1\over 4g^2}F^2-i\bar\theta\{F\tilde F\})]
\end{equation}
where $\{\ \}$ includes the factor $1/32\pi^2$.
It is known that Det factor in the above equation is positive
\cite{vw}. Also note that the $\bar\theta$ term is pure
imaginary. Therefore, using Schwarz inequality, we obtain
the following inequality
\begin{eqnarray}
&e^{-\int d^4x V[\bar\theta]}\le
\int [dA_\mu]\left|\prod_i{\rm Det}(D^\mu\gamma_\mu+m_i)\exp
[-\int d^4x({1\over 4g^2}F^2-i\bar\theta\{F\tilde F\})]\right|
\nonumber\\
&=\int [dA_\mu]\prod_i{\rm Det} (D^\mu\gamma_\mu+m_i)\exp
[-\int d^4x{1\over 4g^2}F^2]=e^{-\int d^4x V[0]}
\end{eqnarray}
which gives
\begin{equation}
V[\bar\theta]\ge V[0].
\end{equation}
Thus $\bar\theta=0$ is the minimum. However, if $\bar\theta$ is
a coupling constant, any $\bar\theta$ can be a good coupling
constant as any $\alpha_{em}$ is allowed theoretically. 
The axion solution interprets $\bar\theta$
as a dynamical field, introducing a kinetic energy term
for for the boson field $\bar\theta$. In this case, we necessarily 
introduce a mass parameter $F_a$, accompanying the axion $a$,
\begin{equation}
\bar\theta={a\over F_a}.
\end{equation}
Then the shape of the potential of $a$ is as shown in Fig.~1.

The hight of the potential is guessed to be of order $\Lambda^4_{
\rm QCD}$. The current algebra calculation gives $(2Z/(1+Z)^2)
f_\pi^2m_\pi^2$. Since the instanton solution gives
$\int d^4x\{F\tilde F\}={\bf Z}$ and $F\tilde F$ appears in the
form given in Eq. (2), {\it $\bar\theta$ is a periodic variable
with period $2\pi$.} Since $\bar\theta$ is a dynamical field,
different $\bar\theta$'s do not describe different theories, but
merely different vacua. Thus, as universe evolves, $\bar\theta$
seeks the minimum of the potential $\bar\theta=0$. This mechanism
explains very elegantly why $\bar\theta$ is so small in our 
universe. The above proof assumed no CP violation except that
from the $\bar\theta$ term, and the weak CP violation shifts
the minimum point very little, $\bar\theta\sim 10^{-17}$~\cite{gr}
which is far below than the bound given by the neutron electric
dipole moment.

An important feature is that $a$ does not have any potential except
that coming from $\bar\theta\{F\tilde F\}$, otherwise the
mechanism does not work. The effect of weak CP violation 
introduces a potential, but as commented above the effect is
very small.

To make $\bar\theta$ dynamical, one must have a mechanism to
introduce a scale $F_a$. Depending on the nature of $a$, one
can classify axion models into three broad categories:\\

\noindent (i) $a$ is the Goldstone boson of a spontaneously broken
chiral $U(1)$ symmetry. The divergence of this $U(1)$ current
must carry the color anomaly $\partial_\mu j^\mu\propto
F\tilde F$ so that $(a/F_a)\{F\tilde F\}$ coupling arises.

%
%

\firstfigfalse
\begin{figure}
$$\epsfig{figure=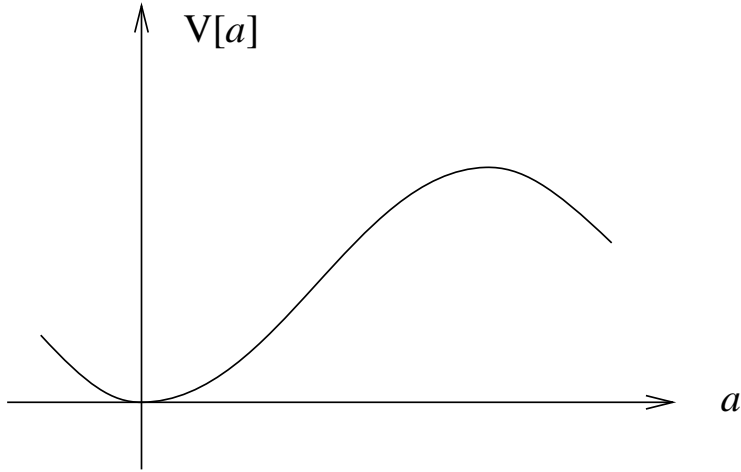,height=2.5in}$$
\caption{A schematic view of the axionic potential with a minimum
at $\theta=0$.}
\label{fig.1}
\end{figure}

\noindent
(ii) $a$ is a {\it fundamental field} in string models.
The scale $F_a$ arises from the compactification. It is
called the model-independent axion~\cite{witten}\\
(iii) $a$ is a composite field. $F_a$ arises at the
confinement scale~\cite{comp}.\\

\subsection{Domain walls}

Because $\bar\theta$ is a periodic variable, the axion potential
looks like as the one shown in Fig.~2. In this example, the origin 
$<a>=0$ is identified with the vacuum $<a>=6\pi F_a$ both of
which are denoted as black dots.
Thus, $<a>=0, 2\pi F_a$, and $4\pi F_a$ are the three degenerate vacua,
distinguished by a black dot, a star, and a triangle.
Since the discrete symmetry of
vacua is spontaneously broken in the evolving universe, there
appear three kinds of domain walls, i.e. $N_{DW}=3$ in our example, 
in the evolving universe. This leads to the so-called axionic 
domain wall problem~\cite{walls}. However, if $N_{DW}=1$, there
seems to be no domain wall problem even if they are formed in
the evolving universe. This is because the string domain wall
network system in the $N_{DW}=1$ model can be erased easily.
A large string attached with a large domain wall dies out due
to punched holes in the wall expands with light velocity
erasing the wall.

If a singlet scalar field develops a
VEV $v$, usually the axion coupling to the gluons has the form
\begin{equation}
{a\over (v/N_{DW})}F_{\mu\nu}^a\tilde F^{a\mu\nu}
\end{equation}
which implies that the coupling is smaller by a factor $N_{DW}$.
Thus the axion mass is larger by a factor $N_{DW}$ if one uses
the vacuum expectation value of the Higgs field. However, if one
uses $F_a$, there does not appear the dependence on $N_{DW}$ as is
evident from the definition of $F_a$ in Eq. (6). One can imagine
a possibility that $10^{14-15}$ GeV scalar vacuum expectation 
value with $N_{DW}\sim 100$ can be consistent with
the cosmological bound. But in this case, of course, $F_a\sim 10^{12}$
GeV.

%
%
\begin{figure}
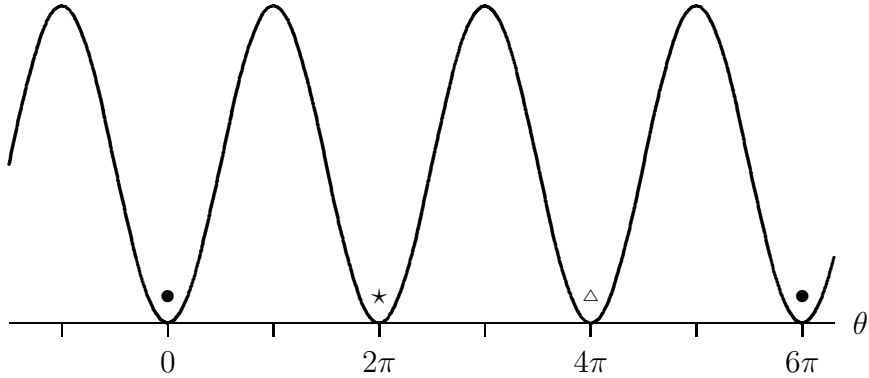

$$\beginpicture
\setcoordinatesystem units <120pt,60pt> point at 0 0
\setplotarea x from -0.500 to 2.100, y from 0.000 to 2.000
\axis bottom 
      ticks out
      width <0.5pt> length <5.0pt>
      withvalues {} {$0$} {} {$2\pi$} {} {$4\pi$} {} {$6\pi$} {} /
      at -0.333 0.000 0.333 0.666 1.000 1.333 1.666 2.000 /
/
\put {$\theta$} <10pt,0pt> at 2.100 0.000
\setplotsymbol ({\normalsize.})
\setsolid
\setquadratic
\plot
-0.500  1.000 -0.450  1.454 -0.400  1.809 -0.350  1.988 -0.300  1.951
-0.250  1.707 -0.200  1.309 -0.150  0.844 -0.100  0.412 -0.050  0.109
 0.000  0.000  0.050  0.109  0.100  0.412  0.150  0.844  0.200  1.309
 0.250  1.707  0.300  1.951  0.350  1.988  0.400  1.809  0.450  1.454
 0.500  1.000  0.550  0.546  0.600  0.191  0.650  0.012  0.700  0.049
 0.750  0.293  0.800  0.691  0.850  1.156  0.900  1.588  0.950  1.891
 1.000  2.000  1.050  1.891  1.100  1.588  1.150  1.156  1.200  0.691
 1.250  0.293  1.300  0.049  1.350  0.012  1.400  0.191  1.450  0.546
 1.500  1.000  1.550  1.454  1.600  1.809  1.650  1.988  1.700  1.951
 1.750  1.707  1.800  1.309  1.850  0.844  1.900  0.412  1.950  0.109
 2.000  0.000  2.050  0.109  2.100  0.412
/
\put {$\bullet$} <0pt,10pt> at 0.000 0.000
\put {$\star$} <0pt,10pt> at 0.667 0.000
\put {\tiny$\triangle$} <0pt,10pt> at 1.333 0.000
\put {$\bullet$} <0pt,10pt> at 2.000 0.000
\endpicture$$
\caption{An axionic potential with $N_{DW}=3$. Here $\theta=a/F_a$.}
\label{fig.2}
\end{figure}

\subsection{Superstring axion}

String models include massless bosons $G_{MN}\ (MN={\rm symmetric}), 
B_{MN}\ (MN={\rm antisymmetric})$, and dilaton $\phi$. Among these, 
$B_{MN}$ is of our interest here. 
Any $D$ dimensional index $M$ can take
$(D-2)$ transverse directions for a massless particle. Therefore,
$B_{MN}$ has $(D-2)(D-3)/2$ physical degrees. In 4 dimensional
Minkowski space time, $B_{\mu\nu}\ \{\mu,\nu=0,1,2,3\}$ has one
physical degree; thus it is a pseudoscalar. The pseudoscalar is
the dual of the field strength of $B_{\mu\nu}$
\begin{equation}
\partial_\sigma a_{MI}\sim\epsilon_{\mu\nu\rho\sigma}H^{\mu\nu\rho}
\end{equation}
where $a_{MI}$ is called the model-independent axion (MIa) in
string models~\cite{witten}. The MIa coupling is universal to
all fermions
\begin{equation}
H_{\mu\nu\rho}\bar\psi\gamma^\mu\gamma^\nu\gamma^\rho\psi\sim
\partial_\mu a_{MI}\bar\psi\gamma^\mu\gamma_5\psi.
\end{equation}
Of course, the coupling of $a_{MI}$ is only of the derivative
form, rendering the nonlinear global symmetry, $a\rightarrow
a+$ (costant). This symmetry is anomalous and the MIa coupling
is universal to all gauge groups \footnote{This comes from the
relation $H=dB+\omega_{3Y}-\omega_{3L}$ where $\omega_{3Y}$
and $\omega_{3L}$ are Yang-Mills and Lorentz Chern-Simons
three forms.}
\begin{equation}
\sim a(F\tilde F+F^\prime\tilde F^\prime+\cdots)
\end{equation} 
Since any superstring model possesses MIa, the axion solution of
the strong CP problem gets a firm theoretical support in string
models. But the axion decay constant is too big in a naive
string models~\cite{choi}. In anomalous $U(1)$ models, however,
the axion decay constant can be lowered~\cite{u1}.

\section{Axion Properties}

Remembering that the axion is a dynamical $\bar\theta$, we can easily
derive its interaction terms. For this, we follow a simple route
of effective field theory.

The simplest axion example is the heavy quark axion~\cite{ksvz}. 
Note that the axion models should provide a pseudoscalar $a$,
coupling to $F\tilde F$. The $a$ is housed in the complex scalar
singlet field $\sigma$. By introducing a heavy quark $Q$, the
following Yukawa coupling is introduced,
\begin{equation}
{\cal L}\propto \sigma\bar Q_RQ_L+{\rm h.c.}
\end{equation}
This model posseses a global Peccei-Quinn symmetry, $Q_L\rightarrow
e^{-i\alpha/2}Q_L, Q_R\rightarrow e^{i\alpha/2}Q_R,
\sigma\rightarrow e^{i\alpha}\sigma$, and $\bar\theta\rightarrow\bar\theta
-\alpha$. The VEV $<\sigma>=F_a/\sqrt{2}$ gives a mass to $Q$, and
produces a Goldstone boson $a$ where $\sigma=(1/\sqrt{2})
(F_a+\rho)e^{ia/F_a}$.
Below the scale $F_a$, the light fields are the gluons and $a$. The
Lagrangian respecting the above symmetry is
\begin{equation}
{\cal L}={1\over 2}(\partial_\mu a)^2 +({\rm derivative\ terms\
of\ }a)+{1\over 32\pi^2}(\theta+{a\over F_a})F_{\mu\nu}^a\tilde
F^{a\mu\nu}.
\end{equation}
Thus, minimally we created a dynamical
variable $\bar\theta=\theta+(a/F_a)$.
It is redefined as $a/F_a$ by shifting the $a$ field. From now on,
$\theta$ implies $\bar\theta$.

Next, let us introduce the known light quarks. As the first
extension, let us consider the up quark condensation in one-flavor 
QCD. The mass term in this theory is given by
\begin{eqnarray}
{\cal L}_{mass} = - m_u \bar{u}_R u_L + {\rm h.c.} 
\end{eqnarray}
Formally, we can assign the following $U(1)$ chiral transformation,
\begin{eqnarray}
u_L &\longrightarrow& e^{i\alpha} u_L\nonumber\\
{u}_R &\longrightarrow& e^{-i\alpha} {u}_R\nonumber\\
m &\longrightarrow& e^{-2i\alpha} m\\
\theta &\longrightarrow& \theta + 2\alpha \nonumber
\end{eqnarray}
Due to the above chiral symmetry, 
we expect the following effective potential
below the chiral symmetry breaking scale $\Lambda_{QCD}$,
\begin{eqnarray}
&V = \frac{1}{2} m_u \Lambda_{QCD}^3 e^{i\theta} - \frac{1}{2}\lambda_1 
  \Lambda_{QCD} v^3 e^{i \frac{\eta}{v} - i\theta}
    -\frac{1}{2}\lambda_2 m_u v^3 e^{i \frac{\eta}{v}}
  +\lambda_3 m_u^2\Lambda_{QCD}^2e^{2i\theta}\nonumber\\
&+\lambda_4{v^6\over \Lambda_{QCD}^2}e^{2i{\eta\over v}-2i\theta}
  +\cdots + \mbox{h.c.}  
\end{eqnarray}
where $\cdots$ is the higher order terms, 
$\lambda$'s are couplings of  order 1, $\langle\bar{u} u\rangle =
v^3 e^{i \eta/v}$, and the QCD scale $\Lambda_{QCD}$ is inserted
to make up the correct dimension. In addition, $e^{\pm i\theta},
e^{\pm 2i\theta}$, etc is multiplied to respect the $U(1)$ symmetry. 
Note that if $m_u \neq 0$ and $\theta$ 
is not a dynamical variable, then the strong CP problem is not solved.
Note that, if $m_u = 0$ then only the $m_u$-independent terms survive,
leading to
\begin{eqnarray}
  V = -{1\over 2}\lambda_1 \Lambda v^3 e^{i \frac{\eta}{v} - i\theta}
  +\lambda_4 {v^6\over\Lambda^2}e^{2i{\eta\over v}-2i\theta}
  +\cdots+\mbox{h.c.} 
\end{eqnarray}
Thus, redefining the $\eta$ field as $\eta^{\prime}$
\begin{eqnarray}
 \eta^{\prime} = \eta - v \theta, 
\end{eqnarray}
the $\theta$ dependence is completely removed from $V$. The $\theta$
parameter is unphysical if a quark is massless. Namely, the massless 
up quark scenario solves the strong CP problem even though it obtains 
a constituent quark mass. The relevance of this solution hinges on
the viability in hadron physics phenomenology \cite{lewt}.
For $m_u\ne 0$, at the minimum $<a>=<\eta>=0$, the mass matrix
is
\begin{eqnarray}
M^2=\left(\begin{array}{cc}
\lambda\Lambda_{QCD}v+\lambda^\prime m_uv &
-\frac{\lambda\Lambda_{QCD}v^2}{F_a}\\
-\frac{\lambda\Lambda_{QCD}v^2}{F_a}&
-\frac{m_u\Lambda_{QCD}^3}{F_a^2}+\frac{\lambda\Lambda_{QCD}v^3}{F_a^2}
\end{array}\right)
\end{eqnarray}
It is easy to calculate determinant of $M^2$
\begin{equation}
{\rm Det}M^2={m\Lambda_{QCD}v\over F_a^2}(\lambda\lambda^\prime)
v^3-\lambda\Lambda_{QCD}^3-\lambda^\prime m\Lambda_{QCD}^2.
\end{equation}
For $F_a\gg {\rm others}$, we obtain $m^2_\eta=(\lambda\Lambda_{QCD}
+\lambda^\prime m)v$. Thus the axion mass is
\begin{equation}
m_a^2={m_u\Lambda_{QCD}\over F_a^2}\left({\lambda\lambda^\prime v^4
\over \lambda\Lambda_{QCD}v+\lambda^\prime m_u v}-\Lambda_{QCD}^2\right)
\end{equation}
which is supposed to be positive. Otherwise we should have chosen
$a=\pi F_a$. This axion mass shows the essential feature: it
is suppressed by $F_a$ and multiplied by $m_u$. The rest is the
condensation parameters. Usually, the condensation parameters are
given in two or three quark flavors.

\subsection{The invisible axion mass}

For two flavors of $u$ and $d$, we can repeat the above argument
with $U(1)_u\times U(1)_d$ symmetry
\begin{eqnarray}
&u_L\rightarrow e^{i\alpha}u_L,\ \  u_R\rightarrow e^{-i\alpha} u_R,\ \ 
d_L\rightarrow e^{i\beta}d_L,\ \  
 d_R\rightarrow e^{-i\beta} d_R\nonumber\\
&m_u\rightarrow e^{-2i\alpha}m_u,\ \  m_d\rightarrow e^{-2i\beta}m_d,
\ \ \theta\rightarrow\theta+2(\alpha+\beta).
\end{eqnarray}  
The effective potential respecting the above symmetry is
\begin{eqnarray}
&V=\h m_um_d\Lambda_{QCD}^2e^{i\theta}-\h {\lambda_1^\prime\over
\Lambda_{QCD}^2}<\bar uu><\bar dd>e^{-i\theta}-\h \lambda_2^\prime
m_u<\bar uu>-\h\lambda_3^\prime m_d<\bar dd>\nonumber\\
&-\h\lambda_4^\prime m_u^*<\bar dd>
e^{-i\theta}-\h\lambda_5^\prime m_d^*<\bar uu>e^{-i\theta}
+\cdots+{\rm h.c.}
\end{eqnarray}
We can diagonalize the $3\times 3$ mass matrix of $a,\pi^0$ and $\eta$
where the phases of $<\bar uu>$ and $<\bar dd>$ are proportional to
${\eta\over F_\pi}+{\pi^0\over F_\pi}$ and ${\eta\over F_\pi}-
{\pi^0\over F_\pi}$, respectively. $\theta$ is proportional to
${a\over F_a}$. The axion mass is
\begin{equation}
m_a\simeq {m_\pi^0F_\pi\over F_a}{\sqrt{Z}\over 1+Z}
\end{equation}
where $Z=m_u/m_d$. The above mass formula is valid for the
very light (or invisible) axion. For the PQWW axion we need
an extra consideration of separating out the longitudinal degree
of the $Z$ boson. Below the chiral symmetry breaking scale the
axion Lagrangian is
\begin{equation}
{\cal L}=\h (\partial_\mu a)^2-\h m_a^2a^2+({\rm interaction\ terms}).
\end{equation}
The interaction terms depend on models.

\subsection{The KSVZ axion}

The KSVZ axion~\cite{ksvz} introduces a heavy quark $Q$,
\begin{equation}
-{\cal L}=f\sigma\bar Q_RQ_L+{\rm h.c.}+m_u\bar u_Ru_L+m_d\bar 
d_Rd_L+\cdots
\end{equation}
where $\sigma$ provides $a$. The light quarks does not transform
under the shift of $a$. At tree level, there does not exists an
axion-electron coupling and it can be induced at one-loop order.

\subsection{The DFSZ axion}

The DFSZ axion~\cite{dfsz} introduces two Higgs doublets
\begin{equation}
-{\cal L}=\lambda\sigma\sigma H_1^*H_2^*+f_u\bar u_Ru_LH_2^0
+f_d\bar d_Rd_LH_1^0+f_e^i\bar e_Re_LH_i^0+{\rm h.c.}
\end{equation}
where $a$ resides mostly in $\sigma$ with a small leakage to
$H_1^0$ and $H_2^0$; phases of $\{H_1^0,H_2^0\}\sim \{\cos\beta,
\sin\beta\}a/F_a$ where $\tan\beta=v_2/v_1$. Depending on models,
$H_1, H_2$, or the third Higgs doublet $H_3$ can couple to 
the electron. For the first two cases, the electron coupling
arises at tree level, $\sim \{\cos\beta,{\rm or\ }
\sin\beta\}(a/F_a)m_e\bar ei\gamma_5e$.

\subsection{The $a-\gamma-\gamma$ coupling}

In view of the possible detection of the cosmic axions in a high-q
cavities immersed in the strong magnetic fields~\cite{detect}, it is
important to know the axion--photon--photon couplings.  More than a
decade ago~\cite{kaplan}, it was calculated, but the current 
citation of the coupling is not accurate. The details of the
KSVZ and DFSZ couplings are given in Ref.~\cite{kim}.
The chiral symmetry breaking at 100 MeV shifts the $a\gamma\gamma$
coupling. Thus the coupling is usually expressed as
\begin{equation}
c_{a\gamma\gamma}=\bar c_{a\gamma\gamma}-{2\over 3}
\cdot {4+Z\over 1+Z}=\bar c_{a\gamma\gamma}-1.92
\end{equation}
where we used $Z=m_u/m_d=0.6$ in the last equation. The $c_{a
\gamma\gamma}$ is the coefficient of $(a/F_a)\{F_{\rm em}
\tilde F_{\rm em}\}$ term. The $\bar c_{a\gamma\gamma}$ is given above
the chiral symmetry breaking scale, and is given by the
Peccei-Quinn symmetry of the theory,
\begin{equation}
\bar c_{a\gamma\gamma}={E\over C},\ {\rm where}\ E={\rm Tr}Q_{\rm em}^2
Q_{PQ},\ \delta_{ab}
C={\rm Tr}\lambda_a\lambda_bQ_{PQ}.
\end{equation}
The normalization is such that  the index $l$ for {\bf 3} and 
{\bf 3}$^*$ is $\h$. The Peccei-Quinn charge is derived from 
the currents obtained from the Lagrangians given in Eqs. (25) and (26),
\begin{eqnarray}
{\rm KSVZ}:&\ J_\mu^{PQ}=\tilde v-\h \bar Q\gamma_\mu\gamma_5Q\\
{\rm DFSZ}:&\ J_\mu^{PQ}=\tilde v+{x^{-1}\over x+x^{-1}}\sum_i
\bar u_i\gamma_\mu\gamma_5u_i+{x\over x+x^{-1}}\sum_i\bar d_i
\gamma_\mu\gamma_5d_i
\end{eqnarray}
where $x=v_2/v_1$ is the ratio of the Higgs doublet VEV's.
It is given in Table 1~\cite{kim}. Here $e_R$ denotes the
electric charge of the representation {\bf R} in units of
the positron charge.

\vskip 0.3cm
\centerline{Table 1. The axion--photon--photon couplings.}
\begin{center}
\begin{tabular}{|cc|cc|}
\hline
\ \ \ \ KSVZ& & DFSZ & \\
$e_R$ & \ \ \ $c_{a\gamma\gamma}$ &  $x$ ($f_e^i$) & \ \ \
$c_{a\gamma\gamma}$\\
\hline
$e_R=0$ &\ \ \ \ --1.92 & any ($i=1$) &\ \ \ \ 0.75\\
$e_3=-\t$&\ \ \ \ --1.25 &1 ($i=2$) &\ \ \ \ --2.17\\
$e_3=\frac{2}{3}$&\ \ \ \ 0.75 &1.5 ($i=2$) &\ \ \ \ --2.56\\
$e_3=1$ &\ \ \ \ 4.08 &60 ($i=2$) &\ \ \ \ --3.17\\
$e_8=1$ &\ \ \ \ 0.75 &1 ($i=3$) &\ \ \ \ --0.25\\
$e_3=-\t,\frac{2}{3}$ &\ \ \ \ --0.25 &1.5 ($i=3$) &\ \ \ \ --0.64\\
 & &60 ($i=3$) &\ \ \ \ --1.25\\
\hline
\end{tabular}
\end{center}
\vskip 0.3cm

The above table cites the couplings in the KSVZ and DFSZ toy
models. In reality, there can be many heavy quarks which carry
nontrivial Peccei-Quinn charges, e.g. as in Ref.~\cite{u1}.
For example, superstring models usually have more than
400 chiral fields. Also, the light quarks are most likely to
carry the Peccei-Quinn charges. Therefore, these effects add
up. In superstring, different models give different values
for $c_{a\gamma\gamma}$. If the standard string model is known,
we can predict the exact value of $c_{a\gamma\gamma}$ in such
a model.

\section{Astrophysical Bounds}

For a sufficiently large $F_a$, axions produced in the stellar
core escape the star easily, which provides an efficient way
of the energy loss mechanism. Comparing this axion emission
process with the standard energy loss mechanism through
neutrino emission gives a bound on $F_a$. The production
cross section is the dominant bottle neck. Thus the
stellar bound is such that enough axions are not produced,
giving a lower bound on $F_a$.  Any axion model has the
Primakoff process of the axion production and nucleon collision
process of the axion production. The DFSZ model has additional
Compton type axion production and similar electron (or positron)
scattering axion production processes. But for $F_a> 10^6$ GeV
which is of our interest here, both the KSVZ and DFSZ models
have similar lower bounds. The astrophysical bounds are reviewed
extensively in the literature~\cite{turner}. For example, from Sun 
one obtains $F_a>2.3\times 10^6(c_a\gamma\gamma/0/75)$ GeV, from red 
giants one obtains $0.9\times 10^8$ GeV, from globular clusters
one obtains $2\times 10^7c_{a\gamma\gamma}$ GeV. For the supernova,
Iwamoto and others studied before the discovery of 
SN1987A~\cite{iwa}. But these pre-SN1987A papers failed to give
a strong lower bound. After SN1987A, many groups obtained the
lower bound of order $10^{9-10}$ GeV~\cite{raffelt}. 
The discrepancy of the numerical studies before and after the
discovery of SN1987A was due to the axion couplings used in
the analyses.
Of course, the correct coupling is of the derivative form
$\partial_\mu a\bar N\gamma^\mu\gamma_5N$ with nucleon $N$.
The correct Goldstone boson nature of the pion is also important
as pointed out in Ref.~\cite{nucl}. In general, this consideration
of the derivative coupling is important at high temperature as
in the supernovae, and gives a lower bound of order~\cite{janka}
\begin{equation}
F_a>10^9\ {\rm GeV}.
\end{equation} 

\section{Cosmic Axion}

The $U(1)$ global symmetry breaking is achieved by a Higgs potential
shown in Fig.~3. The circle at $|<\sigma>|=F_a$ is the axion oscillation
direction. The small perturbation at $|<\sigma>|=F_a$ arises due
to QCD instanton effects. In Fig.~3, there are three degenerate
minima, leading to an $N_{DW}=3$ model. In the evolving universe,
the $U(1)$ breaking at $F_a$ produces axionic strings. At the
chiral symmetry breaking scale of $\sim 100$ MeV, the domain
walls are attached to these axionic strings, which is shown in Fig.~4.
The axion--string and axion--domain--wall system does not die out
quickly if $N_{DW}\ne 1$.

%
%

\begin{figure}
$$\epsfig{file=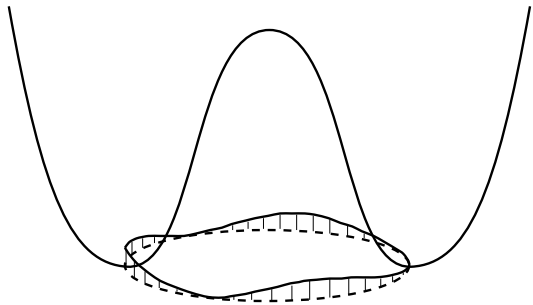,width=70mm}$$
\bigskip
\caption{A potential breaking $U(1)_{PQ}$ with $N_{DW}=3$
arising at the QCD phase transition.}
\end{figure}

%
%

\begin{figure}
$$\epsfig{file=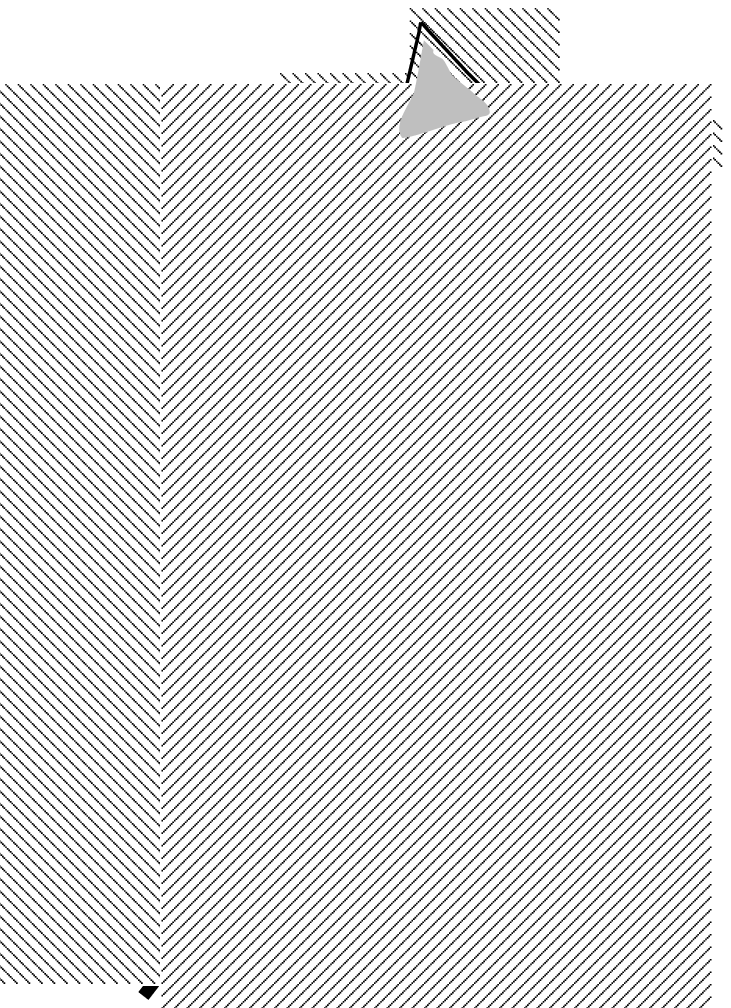,height=75mm}$$
\bigskip
\caption{A schematic view of cosmic string and domain walls with $N_{DW}
=3$.}
\end{figure}

For some time, the reheating temperature after inflation $T_{RH}$
is greater than $F_a$ so that the baryon number generation through
GUT interactions dominates. In this case, we must allow only 
models with $N_{DW}$=1.

But the condition for $N_{DW}=1$ is not necessary if the reheating
temperature after inflation falls below $F_a$. In supergravity
models, the gravitinos which interacts very weakly with observable
sector particles decay so late in cosmic time scale that 
they can dissociate the preciously nucleosynthesized light elements 
unless $T_{RH}<10^{9-10}$ GeV~\cite{ekn}. These arguments favor
a low reheating temperature, presumably below $F_a$. Then the
axionic strings and domain walls are not important at present.

Thus the reliable constraint from cosmology is the cold axion
energy density~\cite{pww}. This arises from the reason of the
invisible axion's extremely feeble coupling so that its lifetime
is many orders larger than the age of the universe. The almost
flat axion potential is felt in the evolving universe when
the Hubble parameter becomes smaller than the axion mass,
$3H<m_a$. This condition is satisfied at the cosmic temperature
$\sim$ 1 GeV. In the inflationary universe, the vacuum value
of $<a>$ is the same in the visible universe. At $T\sim$ 1 GeV, 
$<a>$ begins to roll down the potential hill and continues
the oscillation around $<a>=0$. This motion of the coherent
axion oscillation carries a huge energy density and behaves
like nonrelativistic particles. By now, the estimate of these
cold axion energy density is standard, and one obtains~\cite{cold}
\begin{equation}
\Omega_ah^2\simeq 0.13\times 10^{\pm 0.4}\Lambda_{200}^{-0.7}
f(\theta_1)\left({10^{-5}\ {\rm eV}\over m_a}\right)^{1.18}N^2_{DW}
\end{equation}
where
\begin{equation}
m_a={0.61\times 10^{7}\ {\rm GeV}\over F_a}\ {\rm eV}, 
\end{equation}
$\Lambda_{200}$ is the strong interaction scale in units of 200 MeV,
$\Omega_a$ is the axionic fraction in critical energy density,
$h$ is in units of 100 km/s/Mpc, $\theta_1$ is the initial
misalignment angle taken as $\theta_1\sim \pi/\sqrt{3}$,
and $f(\theta_1)$ is a correction factor of order 1.
The above consideration gives $F_a\le 10^{12}$ GeV not to
close the universe by the cold axions. 

If $N_{DW}=6$, then the axion mass to close the universe is
$0.8\times 10^{-4}$ eV.

Superstring models give $N_{DW}=1$. For $\Omega_a\sim 0.7,
h\simeq 0.65$, the axion mass is $0.5\times 10^{-5}$ eV.

Let us mention the axion string and domain walls in standard
Big Bang or inflation with $T_{RH}>F_a$. In this case, for $N_{DW}
>1$ a complicated axionic string and domain wall network do not
die out easily. Therefore, only $N_{DW}=1$ models are viable.
Even for the $N_{DW}=1$ case, the string-wall system generates
considerable energy. One group asserts that the string-wall system
outweighs the cold axions~\cite{shellard}, while the other group
advocates energy density of axions from walls is 
comparable to or smaller than the cold misalignment 
axions~\cite{harari}. The difference comes
from the different assumptions on the nature of axions radiated
from the axionic walls whether they are cold~\cite{shellard} or 
hot~\cite{harari}. Recent estimate of the axion energy density
from axion walls by Battye and Shellard is dominated by the
axionic string loops~\cite{shellard},  
\begin{equation}
\Omega_a^{string\ loop}=10.7 h^{-2}\Delta
\left[(1+{\alpha\over\kappa})^{3/2}-1\right]\left(
{F_a\over 10^{12}\ {\rm GeV}}\right)^{1.18}
\end{equation}
where $\alpha$ is roughly denoting the loop creation size relative
to the horizon size and $\kappa$ is the back reaction decay rate.
For $\alpha\sim\kappa\sim 0.1$, we have $m_a\ge 100\ \mu$eV or 
$F_a\le 4\times 10^{10}$ GeV. On the other hand, a few years ago
Nagasawa and Kawasaki~\cite{nagasawa} gave a stronger bound 
$F_a\le 10^{10}$ GeV which results from large strings domination 
over the loops. In any case, with $T_{RH}<F_a$ in inflationary
models, this string and wall consideration is irrelevant.  

\section{Axion Searches}

The axion search is really on the invisible axion closing
the universe, for $F_a\sim 10^{12}$ GeV. In this case,
the axion mass to be searched for falls in the several $\mu$eV region.
For models with $N_{DW}=6$, the vacuum expectation value of
the singlet Higgs field is around $10^{13}$ GeV region.

The basic assumption for the cosmic axion detection is that
the cold axions comprise about 70\% of the closure mass density of
the universe. In our galaxy, it is about 0.3 GeV/cm$^3$.
Even though the axion interaction is very weak, the enormous
number density overcome the extremely small conversion rate
of axions to photons. This cosmic
axion detection employs Sikivie's high-q cavities. (Note,
however, the Univ. of Tokyo effort to search for nuclear
M1 transitions~\cite{minowa} which does not
employ Sikivie's detector.)
The photons produced by the cosmic
axion Primakoff process in the strong magnetic field are collected
in the cavity.  There already exist the first round experiments
(the Rochester--Brookhaven--Fermilab (RBF) group
and the University of Florida (UF) group experiments
tried to detect photons collected in this cavity~\cite{rbf}. 
They are shown at the upper right-hand corner in Fig. 5.
The current experiment at LLNL repeats the same type of experiment but
with a bigger cavity~\cite{llnl}. The sensitivity of this new
LLNL experiment as of June, 1997 is also shown in Fig. 5.
Next year, the sensitivity of the LLNL group increases considerably
as shown with a bigger box in Fig. 5.

On the other hand,
the Kyoto group passes the Rydberg atoms in another cavity
where the photons collected in Sikivie's cavity are shone
into the Rydberg atoms; the electrons freed from the Rydberg
atoms by the axion--converted photons are measured~\cite{rydb}. 
 
In Fig.~5, we also show a few model predictions in the KSVZ
and DFSZ models. The DFSZ in Fig. 5 represents the $(d^c,e)$
unification model. 

In Fig.~6, we compare these
data with the predictions from several very light axion models.
As is evident from the figure, it will be difficult to pin
point a toy model even if the cosmic axion is detected.
Most probably, the very light axion would come from the Pecce-Quinn
symmetry breaking where both heavy quarks are the light quarks
carry Peccei-Quinn charges. Superstring models have this
property~\cite{u1}. If a standard superstring model is known
in the future, the axion detection rate will be predicted
in the axion dominated universe without ambiguity.

The detection of the cosmic (or galactic) axions would be a
stunning confirmation of both the particle physics theory (instantons,
invisible axions, etc.) and modern cosmology (galaxy formation,
dark matter, etc.), and would open an window toward a possible
fundamental theory of everything.

%
%

\begin{figure}
$$\beginpicture
\setcoordinatesystem units <240pt,40pt> point at 0 0
\setplotarea x from -6.000 to -4.699, y from -21.50 to -14.00
\inboundscheckon
\linethickness 0.5pt
\axis bottom 
      ticks in
      width <0.5pt> length <5.0pt>
      at -5.699 -5.523 -5.398 -5.301 -5.222 -5.155 -5.097 -5.046 /
      width <1.0pt> length <10.0pt>
      withvalues {$10^{-6}$} {$10^{-5}$} /
      at -6 -5 /
/
\axis top 
      ticks in
      width <1.0pt> length <10.0pt>
      withvalues {$10^{12}$} /
      at -5.2147 /
      width <0.5pt> length <5.0pt>
      withvalues {} {} {} {} {} {} {} {} {} {} {} /
      at -5.993 -5.914 -5.817 -5.692 -5.516 -5.169 -5.118 -5.060
         -4.993 -4.914 -4.817 /
/
\axis top
      ticks out
      width <1.0pt> length <7.0pt>
      withvalues {$4\times10^{12}$} {$4\times10^{11}$} /
      at -5.817 -4.817 /
/
\axis left label {
                  }
      ticks in
      width <0.5pt> length <5.0pt>
      withvalues {} {$10^{-20}$} {} {$10^{-18}$} {} {$10^{-16}$}
                 {} {$10^{-14}$} /
      at -21 -20 -19 -18 -17 -16 -15 -14 /
/
\axis right label {}
      ticks in
      width <0.5pt> length <5.0pt>
      withvalues {} {} {} {} {} {} {} {} /
      at -21 -20 -19 -18 -17 -16 -15 -14 /
/
\put {Axion mass [eV]} <0pt,-15pt> at -5.350 -21.50
\put {$F_a$ [GeV]} <0pt,20pt> at -5.350 -14.00
\setplotsymbol (.)
\setlinear
\put {DARK MATTER BOUND} at -5.611 -20.98
\put {STRING CONT}       at -4.960 -20.98
\setdashes <3pt>
\plot -6.000 -20.45  -5.222 -20.45 /
\plot -4.699 -20.45  -5.222 -20.45 /
\plot -5.222 -21.50  -5.222 -20.45 /
\setshadesymbol <z,z,z,z> ({\tiny.})
\setshadegrid span <4pt>
\vshade -6.000 -21.50 -20.45  -5.222 -21.50 -20.45 /
\setshadegrid span <8pt>
\vshade -5.222 -21.50 -20.45  -4.699 -21.50 -20.45 /
\setdashes <2pt>
\put {\fbox{$e_Q=1$}} <0pt,0pt> [l] at -5.75 -19.33
\plot -6.000 -19.33  -5.750 -19.33 /
\put {\fbox{$e_Q=2/3$}} <0pt,-3pt> [l] at -5.75 -20.00
\plot -6.000 -20.00  -5.750 -20.00 /
\put {\fbox{$e_Q=0$}} [b] at -5.000 -18.85
\plot -6.000 -18.85  -4.699 -18.85 /
\put {\fbox{DFSZ}} [b] at -5.000 -19.66
\plot -6.000 -19.66  -4.699 -19.66 /
\setsolid
\put {LLNL Upgrade} <0pt,5pt> [b] at -5.301 -19.70
\plot -5.890 -14.00   -5.890 -19.75   -4.699 -19.75 /
\put {LLNL Now} <0pt,11pt> [b] at -5.3010 -19.00
\plot -5.865 -14.00  -5.865 -19.00  -4.861 -19.00  -4.861 -14.00 /
\put {\bf Sens.} <-5pt,0pt>   [r] at -5.5744 -17.75
\put {\bf 6/97}  <-5pt,-12pt> [r] at -5.5744 -17.75
\setsolid
\plot -5.564 -14.00  -5.564 -19.00  -5.467 -19.00  -5.467 -14.00 /
\setshadesymbol <z,z,z,z> ({\Large.})
\setshadegrid span <2pt>
\vshade -5.564 -19.00 -14.00  -5.467 -19.00 -14.00 /
\put {\bf UF} <0pt,-5pt> [t] at -5.111 -16.50
\setsolid
\setshadesymbol <z,z,z,z> ({\Large.})
\setshadegrid span <1.5pt>
\plot -5.115 -14.00  -5.115 -16.50  -5.107 -16.50  -5.107 -14.00 /
\vshade -5.121 -16.50 -14.00  -5.101 -16.50 -14.00 /
\plot -5.250 -14.00
-5.250 -16.85  -5.207 -16.85
-5.207 -16.35  -5.173 -16.35
-5.173 -16.30  -5.163 -16.30
-5.163 -14.00 /
\vshade -5.250 -16.85 -14.00  -5.207 -16.85 -14.00 /
\vshade -5.207 -16.40 -14.00  -5.173 -16.40 -14.00 /
\vshade -5.173 -16.35 -14.00  -5.163 -16.35 -14.00 /
\setsolid
\setshadesymbol <z,z,z,z> ({\tiny.})
\setshadegrid span <2pt>
\put {RBF} at -5.045 -14.70
\plot -5.342 -14.00
-5.342 -16.40  -5.294 -16.40
-5.294 -16.00  -5.250 -16.00
/
\plot
-5.163 -16.00  -5.147 -16.00
-5.147 -15.90  -5.135 -15.90
-5.135 -15.80  -5.123 -15.80
-5.123 -15.70  -5.072 -15.70
-5.072 -15.25  -5.040 -15.25
-5.040 -16.00  -5.008 -16.00
-5.008 -16.10  -4.992 -16.10
-4.992 -14.00 /
\plot -4.933 -14.00
-4.933 -15.80  -4.913 -15.80
-4.913 -14.50  -4.861 -14.50
-4.861 -14.92  -4.802 -14.92
-4.802 -15.10  -4.782 -15.10
-4.782 -14.00 /
\vshade -5.342 -16.40 -14.00  -5.294 -16.40 -14.00 /
\vshade -5.294 -16.00 -14.00  -5.250 -16.00 -14.00 /
\vshade -5.163 -16.00 -14.00  -5.147 -16.00 -14.00 /
\vshade -5.147 -15.90 -14.00  -5.135 -15.90 -14.00 /
\vshade -5.135 -15.80 -14.00  -5.123 -15.80 -14.00 /
\vshade -5.123 -15.70 -14.00  -5.072 -15.70 -14.00 /
\vshade -5.072 -15.25 -14.00  -5.040 -15.25 -14.00 /
\vshade -5.040 -16.00 -14.00  -5.008 -16.00 -14.00 /
\vshade -5.008 -16.00 -14.00  -4.992 -16.10 -14.00 /
\vshade -4.933 -15.80 -14.00  -4.913 -15.80 -14.00 /
\vshade -4.913 -14.50 -14.00  -4.861 -14.50 -14.00 /
\vshade -4.861 -14.92 -14.00  -4.802 -14.92 -14.00 /
\vshade -4.802 -15.10 -14.00  -4.782 -15.10 -14.00 /
\endpicture$$
\caption{Cavity experiments performed and planned.}
\label{fig.5}
\end{figure}

%
%

\begin{figure}
$$\epsfig{file=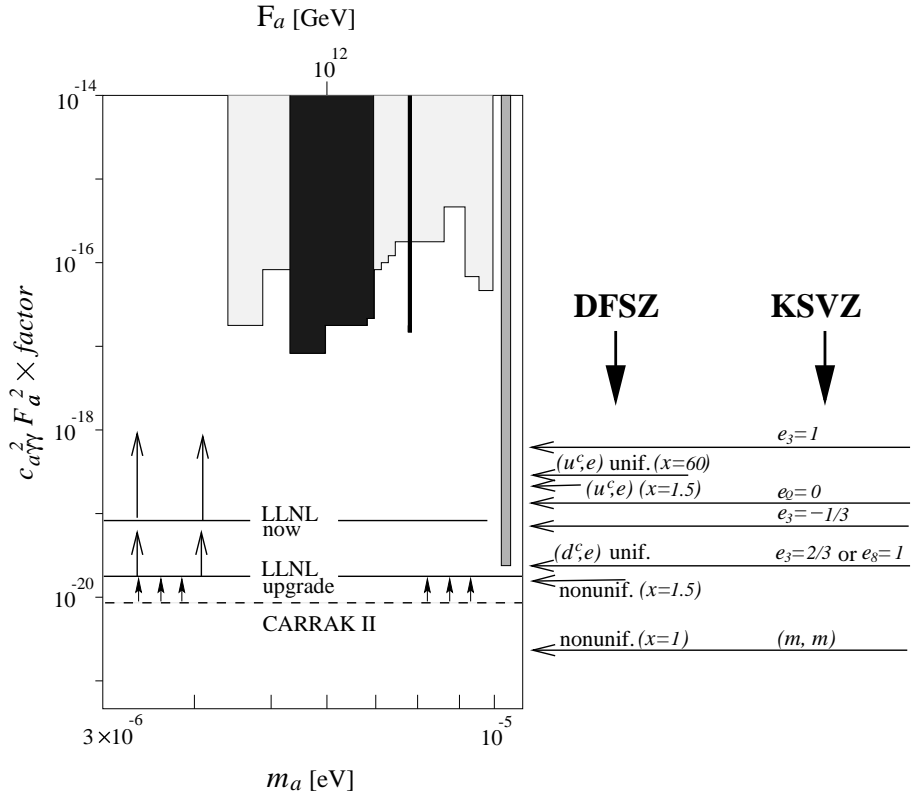,width=120mm}$$
\caption{A part of Fig.~5 compared with several invisible axion models.}
\label{fig.6}
\end{figure}

\vspace{-10mm}

\acknowledgments
I would like to thank Drs. K. Kuroda and M. Kawasaki for
their kind hospitality during the workshop period.
This work is supported in part by the Distinguished Scholar Exchange
Program of Korea Research Foundation and NSF-PHY 92-18167. One of
us (JEK) is also supported in part by the Hoam Foundation.

\end{document}